\documentclass[12pt,epsbox]{article}
\usepackage[nosort]{cite}
\usepackage{color}
\usepackage[dvips]{graphicx,psfrag}
\usepackage{amsmath,amssymb}
\usepackage{bm}
\baselineskip=\normalbaselineskip
\renewcommand{\baselinestretch}{1.34}
\newcommand{\resection}[1]
 {\setcounter{equation}{0}\section{\large{#1}}}

\renewcommand{\thefootnote}{\fnsymbol{footnote}}
\newcommand{\maru}[1]
{{\ooalign{\hfil#1\/\hfil\crcr\raise.167ex\hbox{\mathhexbox20D}}}}

\newcommand{\p}{\partial}

\allowdisplaybreaks[1]

\newcommand{\Lcut}{L_{\rm cut}}

\newcommand{\NN}{\Lambda_{\rm cut}}
\newcommand{\etaAC}{\eta_A^{\rm C}}
\newcommand{\etaAQ}{\eta_A^{\rm Q}}
\newcommand{\bR}{\mbox{\boldmath $R$}}
\newcommand{\vacbra}{\big\langle 0\big|}
\newcommand{\vacket}{\big| 0 \big\rangle}
\begin{document}
\setcounter{page}{0}
\begin{flushright}
\parbox{40mm}{%
KUNS-1888 \\
{\tt hep-th/0312298} \\
December 2003}

\end{flushright}

\vfill

%%%%%%%%%%%%%%%%%%%%%%%%%%%%%%%%%%%%%%%%%%%%%%%%%%%%%
%% Title
%%%%%%%%%%%%%%%%%%%%%%%%%%%%%%%%%%%%%%%%%%%%%%%%%%%%
\begin{center}
{\large{\bf 
A mechanism of the large-scale damping 
in the CMB anisotropy
}}
\end{center}

\vfill

%%%%%%%%%%%%%%%%%%%%%%%%%%%%%%
%% author
%%%%%%%%%%%%%%%%%%%%%%%%%%%%%%
\begin{center}
{\sc Masafumi Fukuma}\footnote%
{E-mail: {\tt fukuma@gauge.scphys.kyoto-u.ac.jp}},  
{\sc Yuji Kono}\footnote%
{E-mail: {\tt kono@gauge.scphys.kyoto-u.ac.jp}} and 
{\sc Akitsugu Miwa}\footnote%
{E-mail: {\tt akitsugu@gauge.scphys.kyoto-u.ac.jp}}  \\[2em]
%$^1$
{\sl Department of Physics, Kyoto University, Kyoto 606-8502, Japan} \\

\end{center}

\vfill
%%%%%%%%%%%%%%%%%%%%%%%%%%%%%%%%%%
% Main
%%%%%%%%%%%%%%%%%%%%%%%%%%%%%%%%%%
\renewcommand{\thefootnote}{\arabic{footnote}}
\setcounter{footnote}{0}
\addtocounter{page}{1}
%%%%%%%%%%%%%%%%%%%%%%%%%%%%%%%%%%%%%%%%%%%%%%%%%%%%%%%%%%%%%%%%
%%%%%%%%%%%%%%%%%%%%%%%%%%%%%%%%%%%%%%%%%%%%%%%%%%%%%%%%%%%%%%%%
%%%%%%%%%%%%%%%%%%%
%% Abstract
%%%%%%%%%%%%%%%%%%%

\begin{center}
{\bf abstract}
\end{center}

\begin{quote}
We present a mechanism through which a certain class of short-distance 
cutoff affects the CMB anisotropies at large angular scales. 
Our analysis is performed in two steps.  
The first is given in an intuitive way, 
using the property of the inflationary universe that 
quantum fluctuations of an inflaton field become classical 
after crossing the Hubble horizon.  
We give a condition for a cutoff 
to yield a damping on large scales, 
and show that the holographic cutoff introduced 
in the preceding paper (hep-th/0307029) 
does satisfy the condition. 
The second analysis is carried out 
by setting an initial condition 
such that each mode of inflaton starts as the vacuum fluctuation 
of the Hamiltonian when being released from the constraint of cutoff. 
The first intuitive discussion is then shown to be correct qualitatively.
\end{quote}
\vfill
%%
%\baselineskip=\normalbaselineskip
\renewcommand{\baselinestretch}{1.4}
%\setlength{\parskip}{0.3\baselineskip}
%%%%%%%%%%%%%%%%%%%%%%%%%%%%%%%%%%%%%%%%%%%%%%%%%%%%%%%%%%%%%%%%
%%%%%%%%%%%%%%%%%%%%%%%%%%%%%%%%%%%%%%%%%%%%%%%%%%%%%%%%%%%%%%%%
\newpage

\resection{Introduction}

Great progress has recently been made in observational cosmology.
In particular, the elaborate measurement of the CMB anisotropy 
\cite{cobe,Bennett:2003bz,Spergel:2003cb,Peiris:2003ff} 
has shown that it can be well explained by the inflationary models 
\cite{Guth:1981,Linde:1981mu,Linde:gd,Linde:1990gz}, 
and has determined many of the relevant 
cosmological parameters from the angular power spectrum 
$C_l$, which is defined through the 
two-point correlators of the temperature fluctuations as
\begin{align}
 \Big\langle \frac{\delta T}{T}(\Omega_1)\,\frac{\delta T}{T}(\Omega_2)
  \Big\rangle =\sum_{l\geq 1} \frac{2l+1}{4\pi}\,
  C_l\,P_l(\cos\theta_{12}). 
\end{align}
Here the bracket $\big\langle~~\big\rangle$ 
is the sample average taken from various pairs of directions 
$\Omega_1$ and $\Omega_2$ on the celestial sphere 
with the fixed angle $\theta_{12}$.

Some discrepancies are known to exist between the observed data 
and the theoretical prediction \cite{Spergel:2003cb}. 
The main contributions to the excess $\chi^2$ originate 
from scales around the first acoustic peak. 
Besides them the observed anisotropies have much smaller values 
at large angular scales. In fact, as for the latter discrepancy, 
the observational data show that $C_l$ is almost proportional 
to $1/l(l+1)$ for small $l$ ($10 \lesssim l<50$), 
consistent with the almost scale invariant power spectrum 
predicted by the inflationary models. 
However, for much smaller $l$ ($l<10$), 
the data show that $C_l$'s are much less than the predicted values
\cite{Spergel:2003cb,Shafieloo:2003gf}.
The main purpose of the present paper 
is to discuss this ``large-scale damping'' 
as a remnant of the Planck-scale physics.

The conventional explanation of this large-scale damping 
is based on the so-called cosmic variance (see, e.g.,%
~\cite{Efstathiou:2003wr,Efstathiou:2003tv,Liddle:cg}). 
In fact, for small $l$, one can take only a few independent samples 
$\big(\!\sim\,2l+1\big)$, and thus the observed values 
can deviate largely from the theoretical mean values. 
However, if this deviation is not simply a statistical deviation, 
then it has a possibility to be a clue to some unknown dynamics 
at the early stage during inflation. 
In fact, as will be reviewed in the next section, 
quantum fluctuations of an inflaton field become classical 
after crossing the Hubble horizon, 
and the CMB anisotropies at large angular scales 
are directly related to those classical fluctuations 
of the inflaton field on large scales. 
Since the larger-scale modes cross the Hubble horizon 
and become classical at earlier times,  
the CMB anisotropies at large angular scales 
are more sensitive to dynamics at the early stage during inflation.

The idea to modify the power spectrum on large scales has a long
history (see, e.g., Refs.\ \cite{Yokoyama:1998rw,Contaldi:2003zv,%
Kawasaki:2003dd}). 
In particular, understanding of trans-Planckian physics has been 
recognized as an important subject.
The effects of trans-Planckian physics on the CMB anisotropies are
discussed in various aspects 
in Refs.\ \cite{Martin:2000xs,Niemeyer:2000eh, Brandenberger:2002nq
,Tsujikawa:2003gh,Huang:2003hw,Chu:2000ww,Cremonini:2003yv
,Easther:2001fz, Tanaka:2000jw,Kaloper:2002cs}.

The most promising candidate that can explain the dynamics 
of quantum gravity at the Planck scale is string theory. 
One of the basic results in string theory 
is the existence of the minimum length scale $l_s$
\cite{Veneziano:1986zf,Yoneya}, 
and spacetime is expected to loose its smooth Riemannian structure 
and to become discrete (or noncommutative) at the Planck scale. 
At the present moment, however,  
we do not have an analytic tool in hand 
with which string dynamics around the Planck scale 
can be dealt with in a definite manner. 
In the present paper, we assume that 
such quantum effects of gravity can be reflected 
simply by introducing a cutoff or a noncommutative scale 
$\Lcut\,\big(=O(l_s)\big)$ into the dynamics of the inflaton field. 
We try to understand the large-scale 
damping by introducing a holographic cutoff to an inflaton
field around the Planck scale.

The possibility that the large-scale damping could be explained 
by a noncommutative nature of spacetime 
was first investigated in Ref.\ \cite{fkm1}.%
\footnote{See Refs.\ \cite{Brandenberger:2002nq,Tsujikawa:2003gh,
Huang:2003hw,Chu:2000ww,Cremonini:2003yv} for inflationary models 
using noncommutative geometry in 
different contexts.}
The main aim of the present article is to complement 
the discussions given there,  
and also to clarify what kind of cutoff should be chosen 
in order to have the large-scale damping in the CMB anisotropy.

This paper is organized as follows. 
In section 2, we explain a mechanism of the large-scale damping, 
using the property of the inflationary universe 
that quantum fluctuations of an inflaton field 
become classical after crossing the Hubble horizon, 
and we show that our previous result given in Ref.\ \cite{fkm1} 
can be naturally understood along this argument. 
However, this process of ``classicalization'' 
should be incorporated automatically 
once a proper initial condition is set  
when an inflaton field starts its quantum fluctuations. 
In section 3, 
we discuss what initial condition should be set 
in the presence of the cutoff introduced in Ref.\ \cite{fkm1} 
and analyze the power spectrum of the inflaton field 
at the exit time of inflation.  
Section 4 is devoted to conclusion and outlook. 

%%%%%%%%%%%%%%%%%%%%%%%%%%%%%%%%%%%%%%%%%%%%%%%%%%%%%%%%%%%%%%%%
%%%%%%%%%%%%%%%%%%%%%%%%%%%%%%%%%%%%%%%%%%%%%%%%%%%%%%%%%%%%%%%%
\resection{A mechanism of the large-scale damping}

We start our discussion with recalling that the currently observed 
CMB power spectrum on the superhorizon scale 
is proportional to the power spectrum of the inflaton field 
at the exit time of inflation. 
We then introduce cutoff in various ways, 
and investigate which kind of cutoff will give rise to 
the large-scale damping. 

%%%%%%%%%%%%%%%%%%%%%%%%%%%%%%%%%%%%%%%%%%%%%%%%%%%%%%%%%%%%%%%%
%%%%%%%%%%%%%%%%%%%%%%%%%%%%%%%%%%%%%%%%%%%%%%%%%%%%%%%%%%%%%%%%
\subsection{Classicalization process 
and the CMB power spectrum in inflationary models (review)}

The flat FRW metric during inflation is given by 
\begin{align}
 ds^2 &=g_{\mu\nu} dx^\mu dx^\nu = a^2(\eta)
 \big( -d\eta^2 + d{\bm x}^2 \big) 
  = a^2(\eta)\,\bigl(-d\eta^2 + dr^2 + r^2 d\Omega^2\bigr),\notag\\
   a(\eta) &= -\frac{1}{H\eta}\quad(\eta<0).
\end{align}
Here $a(\eta)$ is the scale factor and $\eta$ is the conformal time. 
Note that $\eta$ is negative during inflation 
and the exit time of inflation is given by $\eta\rightarrow -0$. 
The inflaton field $\Phi(\eta,{\bm x})$ is decomposed into 
the classical part $\bar{\phi}(\eta)$ 
and the fluctuation $\phi(\eta,{\bm x})$ around it. 
We only consider the Gaussian fluctuation, 
which is realized by taking the part quadratic  in $\phi$ 
after substituting $\Phi=\bar{\phi}+\phi$ into the action 
 $S[\Phi]=\int d^4 x \sqrt{-g}\,
 \bigl( -(1/2)\,g^{\mu\nu}\,\partial_\mu\Phi\,\partial_\nu\Phi
 -V(\Phi) \bigr)$. 
We also assume that the potential $V(\Phi)$ has a plateau 
which is flat enough to ensure that the classical solution 
$\bar{\phi}$ can roll very slowly. 
Thus, in the zeroth order of the slow roll approximation 
we can simply neglect the potential term and obtain 
\begin{align}
 S &= S_{\rm cl} + S_{\rm fluct},\\
 S_{\rm fluct} &= - \frac{1}{2}\int d^4 x\sqrt{-g} \,g^{\mu\nu} 
  \p_\mu \phi \,\p_\nu \phi.
\end{align}
$S_{\rm fluct}$ represents a free scalar field in de Sitter spacetime.

In order to quantize a free scalar field on a curved spacetime, 
one begins with considering a complete set 
of normalized, positive-energy solutions to the Klein-Gordon equation 
\cite{Birrell:ix},  
\begin{align}
 \nabla^2 \psi_A = 0. 
\end{align}
Here $A$ labels the modes with respect to the comoving coordinates,  
and the functions $\psi_A$ are normalized as 
\begin{align}
 (\psi_A, \psi_B) &= \delta_{AB} , & (\psi_A, \psi_B^*) &=0,\\
 (\psi_A^*,\psi_B)&=0, & (\psi_A^*, \psi_B^*) &= -\delta_{AB},
\end{align}
with respect to the symplectic product $(~,~)$ 
that is defined for arbitrary complex functions $f$ and $g$ as 
\begin{align}
 (f, g) &\equiv i \int_\Sigma dS^\mu 
  ( f^* \p_\mu g - \p_\mu f^* \,g)
  =(g, f)^*
  =-(f^*,\,g^*)^*,
\end{align}
where $\Sigma$ is a given spacelike hypersurface, 
and $dS^\mu$ is its three-dimensional volume element. 
Note that $(f,g)$ is independent of the choice of $\Sigma$ 
when $f$ and $g$ satisfy the Klein-Gordon equation. 
Then by introducing the corresponding annihilation and creation 
operators $\{ a_A \}$ and $\{a_A^\dagger\}$ 
with the commutation relations 
\begin{align}
 [a_A, a^\dag_B] &= \delta_{AB}, 
\end{align}
the quantum field $\phi$ is represented as 
\begin{align}
 \phi &= \sum_A ( a_A \,\psi_A + a^\dag _A \,\psi^*_A).
\end{align}
{}For later convenience, 
we denote by $k_A$ the magnitude of the comoving wave number 
of the mode $A$.

The set of modes $\{A\}$ can be chosen as one likes, 
depending on which symmetry is kept manifest. 
We here give a few examples:

\noindent
\underline{(1) $A={\bm k}$}: 
Three-dimensional translational symmetry is kept manifest, 
and 
\begin{align}
 \psi_{{\bm k}}(\eta,{\bm x})&=\frac{H}{\sqrt{2 k^3}}\,
  (1+ik\eta)\,e^{-ik\eta+i{\bm k}\cdot{\bm x}}\quad(k\equiv|{\bm k}|),\\
 \delta_{AA'}&=(2\pi)^3\, \delta^3({\bm k}-{\bm k'}),\quad 
\sum_A =  \int \frac{d^3 \bm{k}}{(2\pi)^3},\\
 k_A&=k.
\end{align}
\noindent
\underline{(2) $A=(k,l,m)$}: 
Rotational symmetry is kept manifest, and 
%the normalized positive-energy functions are given by 
\begin{align}
 \psi_{klm}(\eta,r,\Omega)&=H\sqrt{\frac{2}{k}}\,
  (1+ik\eta)\,e^{-ik\eta}\,j_l(kr)\,Y_{lm}(\Omega)
  \equiv \psi_{kl}(\eta,r)\,Y_{lm}(\Omega), \\
 \delta_{AA'}&= 2\pi\,\delta(k-k')\,\delta_{ll'}\,\delta_{mm'},\quad
\sum_A = \int_0^\infty \frac{dk}{2\pi} \sum_l \sum_m,\\
 k_A&=k,
\end{align}
where $r=|{\bm x}|$ and $\Omega=(\theta,\varphi)$ 
is the angular direction of the comoving coordinate ${\bm x}$.% 
\footnote{
In the above, both of the positive-energy solutions are chosen such that 
they behave as $e^{-ik\eta}$ in the limit $\eta\rightarrow -\infty$. 
It is easy to see that 
the annihilation operators and the positive-energy solutions 
for these two sets of modes
%$a_{\bm k}$ for (1) and $a_{klm}$ for (2), 
are related as 
\begin{align}
 &a_{klm}=\frac{i^l\,k}{2\pi}\int d^2\Omega_{\bm k}\,
  Y^*_{lm}(\Omega_{\bm k})\,a_{\bm k},\\
 &\psi_{\bm k}(\eta,{\bm x})=\frac{2\pi}{k}\,\sum_{l,m}\,i^l\,
  Y^*_{lm}(\Omega_{\bm k})\,\psi_{klm}(\eta,r,\Omega),
\end{align}
where $\Omega_{\bm k}$ is the angular direction of 
the comoving wave number ${\bm k}$. 
}

When rotational invariance is manifest, 
it is often convenient to decompose the field as 
\begin{align}
 \phi(\eta,r,\Omega)=\sum_{l,m}\,\phi_{lm}(\eta,r)\,Y_{lm}(\Omega)
  \quad \Big({\rm or~~}
  \phi_{lm}(\eta,r)=\int d^2\Omega\,Y^*_{lm}(\Omega)\,\phi(\eta,r,\Omega) 
  \Big),
\end{align}
and treat the coefficient 
\begin{align}
 \phi_{lm}(\eta,r)
  =\int_0^\infty \frac{dk}{2\pi}\,\big[ \psi_{kl}(\eta,r)\,a_{klm}
   +(-1)^m\,\psi^*_{kl}(\eta,r)\,a^\dagger_{kl\,-m}\big]
%  =\int_0^\infty dk \,\frac{H}{\sqrt{\pi k}}\,j_l(kr)\,\Big[
%  (1+ik\eta)\,e^{-ik\eta}\,a_{klm}
%  +(-1)^l\,(1-ik\eta)\,e^{ik\eta}\,a_{kl\,-m}\Big]
\end{align}
as a field over a two-dimensional spacetime $(\eta,r)$.
The angular power spectrum of the inflaton field, $C^\phi_l$,  
is then defined by
\begin{align}
 C^\phi_l(\eta,r)&\equiv \vacbra \phi_{lm}^\dagger(\eta,r)
  \,\phi_{lm}(\eta,r)\vacket 
  =\int_0^\infty \frac{dk}{2\pi}\,\big|\psi_{kl}(\eta,r)\big|^2. 
 \label{aps_phi}
\end{align}
Cosmological perturbation theory \cite{Bardeen:kt,Liddle:cg} 
predicts that 
at large angular scales (i.e., for small $l$'s) 
the angular power spectrum $C_l$ 
of the CMB anisotropy observed at present 
is proportional to that of the inflaton at the exit time 
of inflation, 
\begin{align}
 C_l\propto \lim_{\eta\rightarrow -0} C^\phi_l(\eta,r_*). 
 \label{prop}
\end{align}%
\begin{figure}[htbp]
\begin{center}
\resizebox{!}{60mm}{
\input{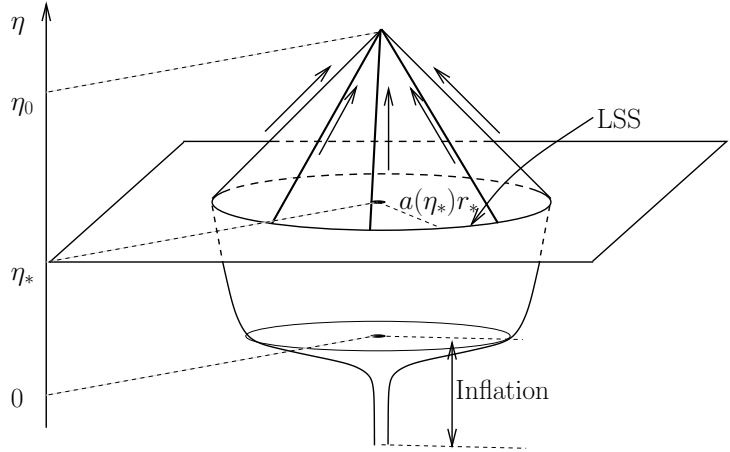}
}
\caption{\footnotesize{
An observer at the present time $\eta = \eta_0$ sees photons from the
LSS as the CMB. 
By setting the comoving radius of the LSS to be $r=r_*$, 
the LSS is traced back to a two-sphere 
with small physical (proper) radius $r_{\rm phys}=a(\eta)\,r$ 
during inflation. 
The moment $\eta=0$ is the exit time of inflation.
}}
\label{LSS}
\end{center}
\end{figure}%
Here $r=r_*$ is the comoving radius of the last scattering surface 
(LSS), which is a two-sphere on the time slice (at $\eta=\eta_*$) 
when the recombination takes place (see Fig.\ \ref{LSS}). 
The relation (\ref{prop}) can be understood as follows. 

\begin{figure}[htbp]
 \begin{center}
  \includegraphics[width=12cm]{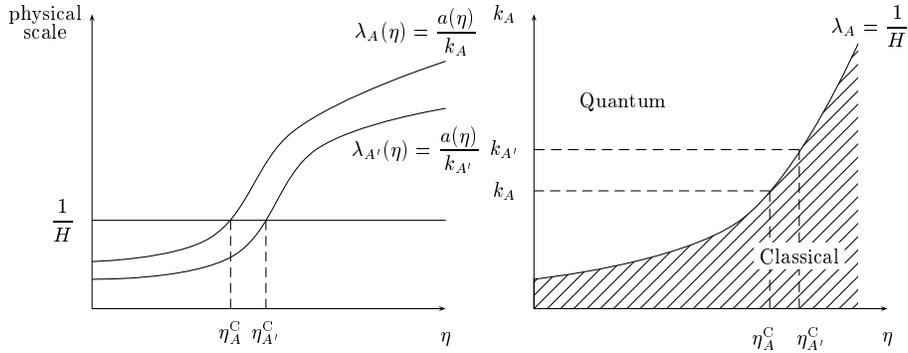}
 \end{center}
 \vspace{-1cm}
 \caption{\footnotesize 
The curve of $\etaAC$ at which the mode $A$ 
crosses the horizon and becomes classical. 
}
\label{class}
\end{figure}

{}First, in the zeroth order of the slow roll approximation 
the Hubble length $1/H$ can be thought to be constant
and the spacetime becomes de Sitter. %during inflation.
In this approximation, once the inflation starts, two points
separated with a distance longer than the Hubble length cannot exchange
information until the end of the inflation.%
%In this case, the Hubble length expresses the physical-distance scale 
%beyond which points cannot be correlated during inflation.%
\footnote{
The Hubble length actually comes to depend on time 
and the spacetime deviates from de Sitter, 
if corrections beyond the zeroth order  
slow roll approximation are taken into account. 
In this case,
one should be careful in analyzing the correlation between two points
 in terms of the Hubble
length.
%because it no longer expresses the causal length. 
See Ref.\ \cite{Davis:2003ad}.
}
Thus, a mode $A$ can fluctuate quantum mechanically 
only when its  physical wave length $\lambda_A$ is shorter than 
the Hubble length. 
However, in the expanding universe, 
the physical wave length of the mode $A$ 
is a monotonically increasing function of time 
as $\lambda_A(\eta)=a(\eta)/k_A$, 
while the Hubble length is constant in time during inflation. 
Thus, as the physical wave length increases, 
the mode $A$ gradually looses its quantum nature 
and eventually becomes purely classical 
at the exit time of inflation $\eta\rightarrow -0$ 
(for a detailed analysis, see, e.g., Ref.\ \cite{Albrecht:1992kf}).  
This process is sometimes called the classicalization.  
The characteristic moment $\eta_{A}^{\rm C}$ 
at which the crossover from quantum to classical physics occurs 
for the mode $A$, can be evaluated by setting $\lambda_A=1/H$ 
(crossing the Hubble horizon) 
and is found to be $\eta_A^{\rm C}=-1/k_A$ (see Fig.\ \ref{class}).
We thus can schematically say that 
\begin{align}
 ~~%{\rm if~} 
  \eta < \etaAC &\longrightarrow 
  \text{the mode $A$ can fluctuate quantum mechanically,}\notag\\
 ~~%{\rm if~} 
  \eta > \etaAC &\longrightarrow 
  \text{the mode $A$ cannot fluctuate quantum mechanically.}\notag
\end{align}
%%%%%%%%%%
%%%%%%%%%%
Note that a mode of smaller $k_A$ crosses the horizon 
at an earlier time.

After the ending of inflation, 
one can apply the linear perturbation theory,
regarding $\phi_{lm}(\eta,r)$ in Eq.\ (\ref{aps_phi}) 
simply as a classical perturbation. 
One can show that the CMB anisotropies on large scales are 
proportional to the perturbation of gravitational potential 
on the LSS (Sachs-Wolfe effect 
\cite{Sachs,Liddle:cg}).
The latter in turn can be linearly related to the classical 
fluctuations of the inflaton at the exit time of inflation 
by using the conservation law in the cosmological perturbation theory 
which holds on the superhorizon scale with adiabatic perturbation 
\cite{Bardeen:kt,Liddle:cg}. 
Putting it all together, 
the angular mode $c_{lm}$ of the CMB anisotropy 
\begin{align}
 \frac{\delta T}{T}(\Omega)
  =\sum_{l\geq 1}\sum_{m=-l}^l\,c_{lm}\,Y_{lm}(\Omega)
\end{align}
is proportional to $\phi_{lm}(\eta\!=\!-0,\,r\!=\!r_*)$,  
and thus, the angular power spectrum $C_l$ of 
the CMB anisotropy observed at the present time is proportional 
to that of the inflaton field at the exit time of inflation:
\begin{align}
 C_l =\big\langle |c_{lm}|^2 \big\rangle
  \propto \lim_{\eta\rightarrow -0}
  \big\langle |\phi_{lm}(\eta,r_*)|^2 \big\rangle
  = \lim_{\eta\rightarrow -0} C^\phi_l(\eta,r_*).
\end{align}
In the zeroth order of the slow roll approximation, 
we have 
$\lim_{\eta\rightarrow-0}\psi_{kl}(\eta,r_*)
 =H\sqrt{2/k}\,j_l(kr_*)$, 
so that 
\begin{align}
 C_l \propto
  C_l^\phi(0,r_*) =
  \frac{H^2}{\pi}\,\int_0^\infty\frac{dk}{k}\,\big(j_l(kr_*)\big)^2
  \propto \frac{1}{l(l+1)}. 
%  &= \left | \frac{H}{5 \dot{\bar{\phi}}} \right|^2 
%  \langle \phi^\ast_{lm}\phi_{lm} \rangle  
%  = 4\pi\left | \frac{H}{5 \dot{\bar{\phi}}} \right|^2 
%  \int_0^\infty \frac{dk}{k} \bigl(j_l(kr_*)\bigr)^2 P_\phi (k),
 \label{base}
\end{align}
%The leading order of the slow roll approximation 
%gives the scale invariant power spectrum at the exit time of inflation%
%\footnote{
%This result will be derived in the next subsection 
%in a more general situation.
%} 
%\begin{align}
% P_\phi(k ) \equiv \lim_{\eta\rightarrow -0} P_{\phi}(\eta,\bm k)
%  = \left( \frac{H}{2\pi}\right)^2,  
%\end{align}
%which leads to the angular power spectrum
%\begin{align}
% C_l \propto \int_0^\infty \frac{dk}{k}\,\bigl(j_l(kr_*)\bigr)^2 
%  \propto 1/l(l+1).
%\end{align}
This prediction from inflationary models agrees quite well 
with the observational data at large angular scales.

%%%%%%%%%%%%%%%%%%%%%%%%%%%%%%%%%%%%%%%%%%%%%%%%%%%%%%%%%%%%%%%%%%%
%%%%%%%%%%%%%%%%%%%%%%%%%%%%%%%%%%%%%%%%%%%%%%%%%%%%%%%%%%%%%%%%%%%
\subsection{Time-dependent cutoff and a criterion for 
the large-scale damping}

In this subsection, we discuss how the previous analysis 
should be modified when various types of cutoff are introduced. 
One may think that nothing would change on large scales 
since the short-distance cutoff usually does not play 
important roles on large scales in local quantum field theories. 
However, in the expanding universe, 
the cutoff can be a time-dependent function for the comoving modes. 
This fact, combined with the classicalization process, 
proves to give rise to large suppression 
on the large-scale modes when a particular form of cutoff is chosen.

%%%%%%%%%%
\begin{figure}[htbp]
 \begin{center}
  \includegraphics[width=7cm]{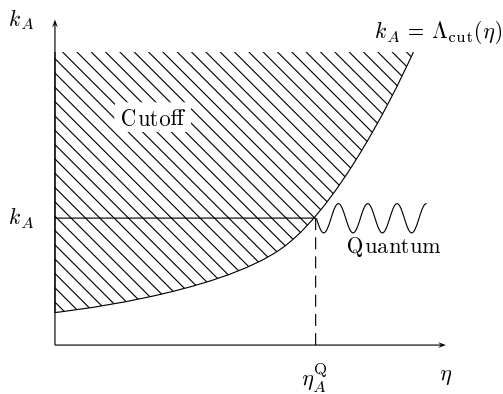}
 \end{center}
 \vspace{-1cm}
 \caption{\footnotesize 
The time-dependent comoving cutoff. 
In the shadowed region, quantum fluctuations are prohibited. 
The mode $A$ starts its quantum fluctuation at the moment $\eta=\etaAQ$. 
}
\label{cutoff}
\end{figure}
%%%%%%%%%%
Suppose that we introduce a short-distance cutoff $\Lcut$ on the
physical distance scale such that it is constant in time. 
Then there appears another moment which will be important 
in the following discussion.
In fact, as we see below, this leads to a 
cutoff $\NN(\eta)$ on comoving modes, which is 
generically a monotonically increasing function of time 
in the expanding universe (see Fig.\ \ref{cutoff}), 
and whose dependence on the short-distance cutoff $\Lcut$ 
strongly depends on the way of introducing $\Lcut$ into a system. 
A comoving mode $A$ can exist as a quantum fluctuation 
only when the inequality 
$k_A \leq\NN(\eta)$ is satisfied. 
We thus can introduce the moment $\etaAQ$ 
at which the mode $A$ starts its quantum fluctuation 
upon being released from the constraint of the cutoff. 
In a nutshell,  
\begin{align}
 ~~%{\rm if~}
  \eta  &< \etaAQ \longrightarrow 
  \text{the mode $A$ is prohibited to exist as a quantum fluctuation,}
  \notag\\
 ~~%{\rm if~}
  \eta &> \etaAQ \longrightarrow 
  \text{the mode $A$ is allowed to fluctuate quantum mechanically.}
  \notag
\end{align}

In order to calculate the power spectrum we need to fix 
the initial condition at $\eta = \etaAQ$ for each mode $A$.
Although we have no established one,
we know which form the initial conditions should take 
for the modes in the two extreme cases, 
$\etaAQ \ll \etaAC$ and $\etaAC \ll \etaAQ$. 
In the former case, the physical length scales of such modes $A$ 
upon being created are much smaller than the Hubble length. 
Therefore, as is usual, the Bunch-Davies vacuum 
\cite{Birrell:ix} is favored 
and the contribution of these modes to the CMB anisotropy
should be the same as Eq.\ (\ref{base}).
On the other hand, as for the latter case $\etaAC \ll \etaAQ$, 
we first recall that the anisotropies of the CMB
observed at the
present time correspond to the classical values (or the `fossils') of
the quantum fluctuations of the inflaton field (see the discussion in
the previous subsection).
We also note that  a mode $A$ with $\etaAC \ll \etaAQ$ 
must be already classical well before its quantum fluctuation starts.
Since no fossil can exist if the mode
does not have a life of quantum fluctuation, such a mode with 
$\etaAC \ll \etaAQ$ has no classical amplitude and thus does
not contribute to the CMB anisotropy.

In between these two limits, we do not a priori
know how modes contribute to the CMB
anisotropies without setting definite initial conditions
on quantum fluctuations.
We consider this problem in detail in section 3.
Here, we instead set the following ansatz, expecting from the above
argument that
those modes with $\etaAQ < \etaAC$ contribute to the CMB
anisotropies more than those with 
$\etaAC < \etaAQ$ do:

\noindent\underline{\textbf{Ansatz}}
\begin{align}
\left\{ 
\begin{array}{ccl}
 \, \bullet\; \eta^{\rm C}_A < \eta^{\rm Q}_A &\longrightarrow&~ 
  \text{the mode $A$ does not contribute to the CMB anisotropy,}\\
 \, \bullet\; \eta^{\rm Q}_A < \eta^{\rm C}_A &\longrightarrow&~  
  \text{the mode $A$ contributes to the CMB anisotropy as in}\\
 & &~  \text{the absence of cutoff.}
\end{array}\right. \label{ansatz}
\end{align}

Whether the cutoff gives rise to a damping of the CMB anisotropy 
at large distance scales or at short distance scales, 
will depend on the way of introducing cutoff. 
%%%%%%%%%%
\begin{figure}[htbp]
 \begin{center}
  \includegraphics[width=7cm]{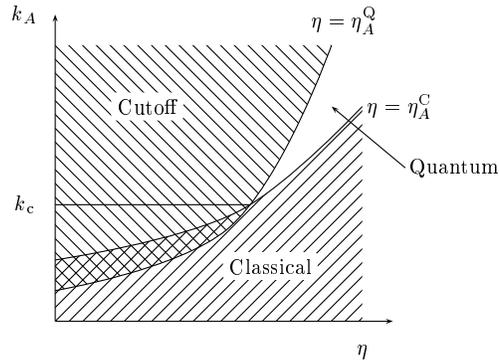}
 \end{center}
 \vspace{-1cm}
 \caption{\footnotesize 
Case in which a damping occurs on scales larger than $1/k_{\rm c}$. 
}
\label{largescale}
\end{figure}
%%%%%%%%%%
If $\etaAQ$ and $\etaAC$ 
behave as in Fig.\ \ref{largescale}, 
then larger-scale modes do not contribute 
and the CMB anisotropy has a damping on large scales. 
%%%%%%%%%%
\begin{figure}[htbp]
 \begin{center}
  \includegraphics[width=7cm]{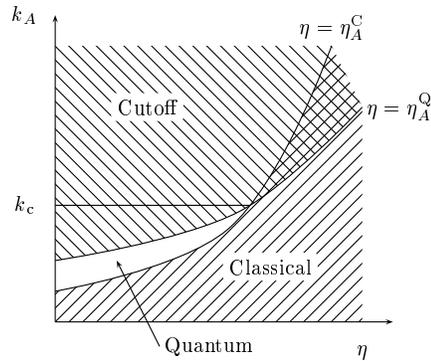}
 \end{center}
 \vspace{-1cm}
 \caption{\footnotesize 
Case in which a damping occurs on scales smaller than $1/k_{\rm c}$. 
}
\label{smallscale}
\end{figure}
%%%%%%%%%%
On the other hand, if their behaviors are as in Fig.\ \ref{smallscale}, 
then one would observe a damping on small scales. 
%%%%%%%%%%
\begin{figure}[htbp]
 \begin{center}
  \includegraphics[width=7cm]{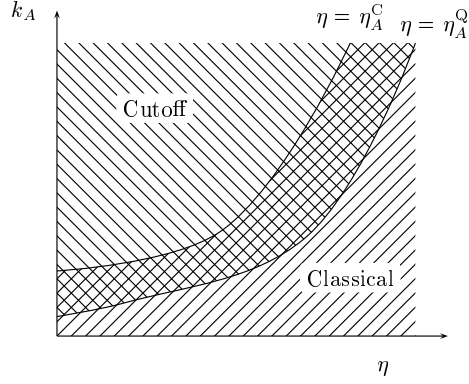}
 \end{center}
 \vspace{-1cm}
 \caption{\footnotesize 
Case in which no quantum fluctuations occur during the whole period. 
No CMB anisotropies will be made. 
}
\label{nocmb}
\end{figure}
%%%%%%%%%%
%%%%%%%%%%
\begin{figure}[htbp]
\begin{center}
  \includegraphics[width=7cm]{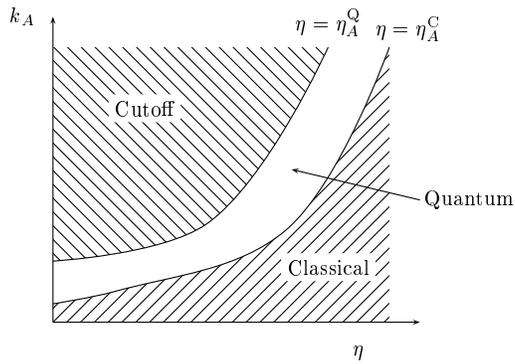}
 \end{center}
 \vspace{-1cm}
 \caption{\footnotesize 
Case in which all the quantum fluctuations are allowed 
to exist during inflation. 
This will give almost the same result with that in the absence of cutoff.
}
\label{nodamping}
\end{figure}
%%%%%%%%%%
In the case of Fig.\ \ref{nocmb}, 
there are no quantum modes and will be no 
classical fluctuations, 
while in the case of Fig.\ \ref{nodamping}, 
all the quantum modes are allowed to exist, 
and one would not see a major deviation from 
the case in the absence of cutoff. 

%%%%%%%%%%%%%%%%%%%%%%%%%%%%%%%%%%%%%%%%%%%%%%%%%%%%%%%%%%%%%%%%%%%
%%%%%%%%%%%%%%%%%%%%%%%%%%%%%%%%%%%%%%%%%%%%%%%%%%%%%%%%%%%%%%%%%%%
\subsection{A few examples}

We give a few examples on the way of introducing cutoff. 
In the following, 
we expand the inflaton field $\phi(\eta,r,\Omega)$ 
with respect to angular coordinates as 
\begin{align}
 \phi(\eta,r,\Omega)
  =\sum_{l\geq 0}\,\sum_{m=-l}^{l}\,\int_0^\infty \frac{dk}{2\pi}\,
  \phi_{k l m}(\eta)\,j_l(kr)\,Y_{lm}(\Omega), 
 \label{angular_decomposition}
\end{align}
and label the mode as $A=(k,l,m)$. 
In Eq.\ (\ref{angular_decomposition}), 
$k$ denotes the comoving wave number in the radial direction 
and $(l,m)$ represents the multipole in the angular directions.  

%%%%%%%%%%%%%%%%%%%%%%%%%%%%%%%%%%%%%%%%%%%%%%%%%%%%%%%%%%%%%%%%%%%
\subsubsection{Homogeneous cutoff on the physical wave length}

\noindent 
We first consider the case in which 
a time-independent short-distance cutoff $\Lcut$ 
is put on the physical wave length. 
That is, we require that the mode $A$ can exist 
only when the following inequality holds:
\begin{align}
 \frac{a(\eta)}{k } \geq \Lcut \label{cutone}. 
\end{align}
This can be imagined as that the three-dimensional space is a lattice 
with the spacing $\Lcut$. 

By setting 
\begin{align}
 \frac{a(\etaAQ)}{k} &= - \frac{1}{ k H \etaAQ} 
  \equiv \Lcut,
\end{align}
the moment $\etaAQ$ at which the mode starts 
its quantum fluctuation is found to be 
\begin{align}
 \etaAQ &= - \frac{1}{k \Lcut H}. \label{etaaq} 
\end{align}
On the other hand, 
the moment $\etaAC$ at which the mode crosses the horizon 
and becomes classical, is determined by setting 
\begin{align}
 \lambda_A=\frac{a(\etaAC)}{k} 
  &=- \frac{1}{ k H \etaAC}   \equiv\frac{1}{H},
\end{align}
or
\begin{align}
 \etaAC &= - \frac{1}{k}.
\end{align}
%%%%%%%%%%
\begin{figure}[htbp]
 \begin{center}
  \includegraphics[width=7cm]{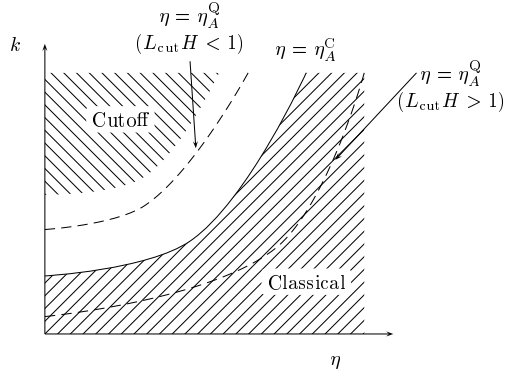}
 \end{center}
 \vspace{-0.7cm}
 \caption{\footnotesize 
Homogeneous cutoff on the physical wave length. 
The figure has the same pattern with that of 
Fig.\ \ref{nocmb} or Fig.\ \ref{nodamping} 
depending on the value of $\Lcut H$.
}
\label{wave}
\end{figure}%
%%%%%%%%%%
$\etaAC$ and $\etaAQ$ are depicted in Fig.\ \ref{wave}, 
where the vertical axis represents $k_A=k$. 
This figure corresponds to Fig.\ \ref{nocmb} or Fig.\ \ref{nodamping}, 
depending on the value of $\Lcut H$ ($>1$ or $<1$), 
and thus we conclude that this way of introducing cutoff 
does not yield a particular damping in the CMB anisotropy. 

%%%%%%%%%%%%%%%%%%%%%%%%%%%%%%%%%%%%%%%%%%%%%%%%%%%%%%%%%%%%%%%%%%%
\subsubsection{Fuzzy sphere}

We next consider the cutoff \cite{fkm1}
with which the two-sphere for each $(\eta,r)$ 
is replaced by a fuzzy sphere, 
such that only one-bit degrees of freedom 
can reside on each area $\Lcut^2$
\cite{hppe1982}. 
\begin{figure}
\begin{center}
\resizebox{!}{40mm}{
 \input{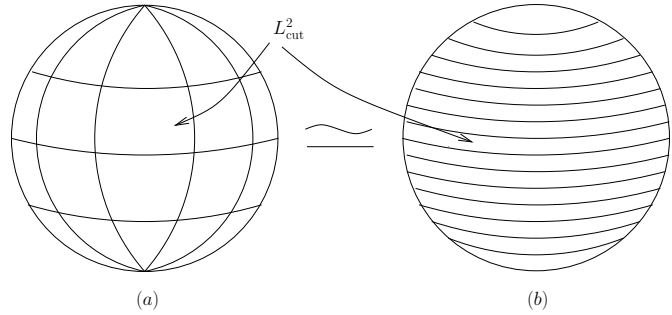}
}
\end{center}
\caption{\footnotesize 
Fuzzy sphere used as a cutoff. 
(a) The fuzzy sphere consists of $(N+1)$ points, 
each of which is represented as a two-dimensional region of area $\Lcut^2$.
(b) By using the noncommutative algebra for the spatial coordinates 
$\hat{\bm x}=(\hat{x}^1,\hat{x}^2,\hat{x}^3)$ 
with the commutation relations 
$\big[\hat{x}^i,\hat{x}^j\big]=\big(2ri/\sqrt{N(N+2)}\big)\,
\epsilon^{ijk}\,\hat{x}^k$, 
the fuzzy sphere of radius $r$ is represented 
as the set of $(N+1)$ looped strips of equal area $\Lcut^2$ when a
particular direction (say, $\hat{x}^3$) is diagonalized 
\cite{hppe1982}. }
\label{lcut}
\end{figure}
Then the fuzzy sphere consists of finitely many points 
(see Fig.\ \ref{lcut}), 
the number of which is given by 
\begin{align}
 \frac{4\pi a^2(\eta)\,r^2}{\Lcut^2}
  =\frac{4\pi r^2}{\Lcut^2 H^2 \eta^2 }
  \equiv N(\eta,r)+1. 
\end{align}
This in turn gives an upper bound on $l$ as 
(see, e.g., \cite{fkm1})
\begin{align}
 l \leq N(\eta,r) = \frac{4\pi r^2}{\Lcut^2 H^2 \eta^2 }-1.   
\label{lmax}
\end{align}
If we consider the orbit of the LSS  
with the fixed comoving radius $r=r_*$, 
the moment $\etaAQ$ 
%at which the mode $A=(k,l,m)$ starts its quantum fluctuation 
is given by solving the equation 
\begin{align}
 l &= N(\etaAQ, r_*) 
  = \frac{4\pi r^2_*}{\Lcut^2H^2 \bigl(\etaAQ\bigr)^2} -1 ,
\end{align}
and thus found to be 
\begin{align}
 \etaAQ &= -\frac{r_*}{\Lcut H} \sqrt{ \frac{4\pi}{l + 1}} \label{etal}. 
\end{align}
The moment $\etaAC$ 
%at which the mode crosses the horizon and becomes classical, 
is again given by
\begin{align}
 \etaAC &= - \frac{1}{k}. 
\end{align}

%%%%%%%%%%
\begin{figure}[htbp]
 \begin{center}
  \includegraphics[width=7cm]{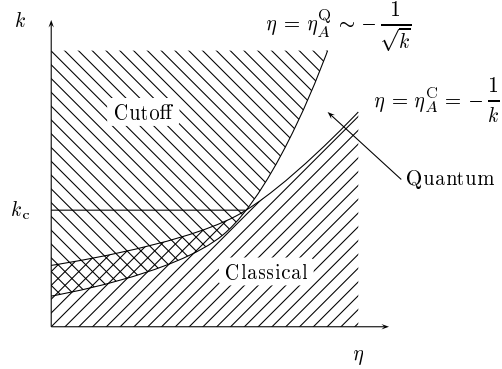}
 \end{center}
 \vspace{-1cm}
 \caption{\footnotesize 
Cutoff with fuzzy sphere. 
This has the same pattern with that of Fig.\ \ref{largescale} 
and a damping will occur at large distance scales.
}
\label{fuzzy}
\end{figure}
%%%%%%%%%%
These two moments $\etaAQ$ and $\etaAC$ 
can be compared 
if one notes that $k$ can be identified with $\pi(l+1)/2r_*$ 
$\big({\textrm{i.e.,}}\,\etaAQ\sim-\pi\sqrt{2 r_*}/(\Lcut H \sqrt{k})\big)$
since the spherical Bessel function $j_l(kr_*)$ 
has a sharp peak at the value $kr_*\sim \pi(l+1)/2$. 
The result is given in Fig.\ \ref{fuzzy}.
This figure has the same pattern with that of Fig.\ \ref{largescale} 
and thus leads to a damping at large distance scales. 

This line of argument was given in Ref.\ \cite{fkm1}. 
There the analysis was made by setting the ansatz \eqref{ansatz} that 
once a quantum mode has a period of quantum fluctuation,  
the classical value has the magnitude 
same with that of the case in the absence of cutoff. 
This ansatz is equivalent to the prescription 
that the normalized positive-energy solution 
in the absence of cutoff,
\begin{align}
 \psi^{(0)}_{A=(k,l,m)}(\eta,r,\Omega)
  =H\sqrt{\frac{2}{k}}\,\big(1+ik\eta\big)\,j_l(kr_*)\,
  e^{-ik\eta}\,Y_{lm}(\Omega), 
\end{align}
is replaced by 
\begin{align}
 \psi_{A}(\eta,r,\Omega)
  &\equiv\theta\big(\etaAC-\etaAQ\big)\,
  \psi^{(0)}_{A}(\eta,r,\Omega)\nonumber\\
 &=\theta\Big(k-\frac{\alpha_l}{r_*}\Big)\,
  \psi^{(0)}_{A}(\eta,r,\Omega)\quad
  \biggl(\alpha_l\equiv \Lcut H\,\sqrt{\frac{l+1}{4\pi}}\biggr)
\end{align}
in the presence of the noncommutativity. 
This gives an IR cutoff on the integration over the comoving wave number, 
and the angular power spectrum of inflaton is found to be
\begin{align}
 \vacbra \phi^\dagger_{lm}(\eta,r_*)\, \phi_{lm}(\eta,r_*)
  \vacket\Big|_{\eta\rightarrow -0}
  =\frac{H^2}{\pi}\int_{\alpha_l/r_*}^\infty \frac{dk}{k}\,
  \big( j_l(kr_*)\big)^2. 
\end{align} 
If we further introduce the spectral index $n~(\sim1)$, 
then the angular power spectrum becomes
\begin{align}
 \vacbra \phi^\dagger_{lm}(\eta,r_*)\, \phi_{lm}(\eta,r_*)
  \vacket\Big|_{\eta\rightarrow -0}
  =\frac{H^2}{\pi}\int_{\alpha_l/r_*}^\infty \frac{dk}{k}\,
  k^{n-1}\,\big( j_l(kr_*)\big)^2, 
\end{align} 
which gives the angular power spectrum of the CMB anisotropy as 
\begin{align}
 C_l=(1-\beta_l)\,C_l^{(0)}, 
\end{align}
where the $C_l^{(0)}$ is the value when noncommutativity is absent 
during inflation. The damping factor $(1-\beta_l)$ is given by 
\begin{align}
 \beta_l=\frac{4}{\sqrt{\pi}}\,
  \frac{\Gamma\big((4-n)/2\big)\Gamma\big(l+(5-n)/2\big)}
  {\Gamma\big((3-n)/2\big)\Gamma\big(l+(n-1)/2\big)}\,
  \int_0^{\alpha_l}dx\,x^{n-2}\,\big(j_l(x)\big)^2, 
\end{align}
%%%%%%%%%%
\begin{figure}[htbp]
 \begin{center}
  \rotatebox{270}{
  \resizebox{!}{100mm}{\includegraphics{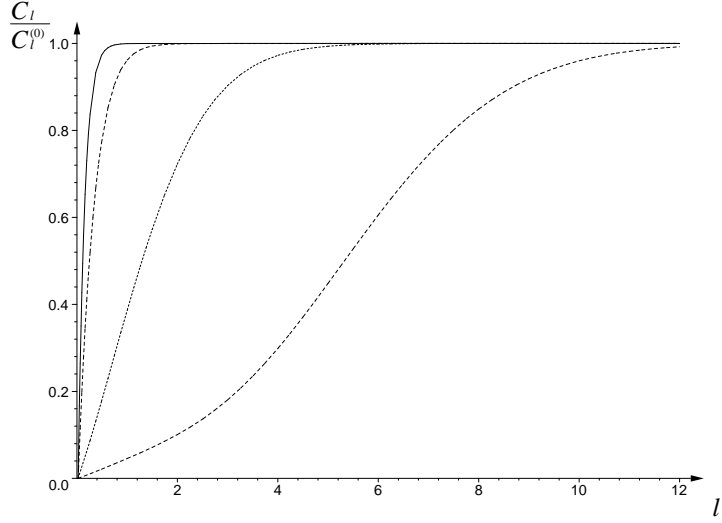}}}
 \end{center}
 \caption{\footnotesize 
  Damping factor for the model using fuzzy sphere. 
  The horizontal axis represents the 
  multipole $l$, while the vertical axis is
  $C_l/C_l^{(0)}=1-\beta_l$. 
%  We have normalized the constant such that it takes one
%  when all the noncommutativity is removed. 
  Here the spectral index is set to be $n=0.95$. 
  $\Lcut H$ is 0.1, 1, 5, 10 from top to bottom \cite{fkm1}.
}
\label{N=0.95}
\end{figure}
%%%%%%%%%%
and is depicted in Fig.\ \ref{N=0.95} 
with $n=0.95$. 

%%%%%%%%%%%%%%%%%%%%%%%%%%%%%%%%%%%%%%%%%%%%%%%%%%%%%%%%%%%%%%%%%%%
\subsection{Connection with the existing literature}
So far we have discussed the effects of cutoff on the CMB anisotropies
by comparing the two moments $\etaAC$ and $\etaAQ$.
In the existing literature, discussions are usually given by
introducing an  energy scale $M_c$, 
where the modes are supposed to be created 
(see 
Refs.\ \cite{Martin:2000xs,Niemeyer:2000eh,%
Brandenberger:2002nq,Tsujikawa:2003gh,%
Huang:2003hw,Chu:2000ww,Cremonini:2003yv,%
Easther:2001fz,Tanaka:2000jw,Kaloper:2002cs}).
Then the only other scale in the problem is 
the energy scale of inflation, 
$H$, and thus the effect of the new scale is expressed 
by the dimensionless ratio $H/M_c$.
In the literature it is commonly assumed that $M_c$ is 
almost constant in time. In this subsection we argue 
that the large-scale damping can occur if $M_c$ is a 
monotonically increasing function of time. 
We further show that the cutoff based on the fuzzy sphere (subsection
2.3.2) actually leads to $M_c$ of this kind.

With the use of $M_c$, 
the moment $\etaAQ$ at which the quantum mode $A$ is created 
is characterized by the equation 
$a(\etaAQ)/k_A=1/M_c(\etaAQ)$,  
where we have assumed the possible dependence of $M_c$ on time. 
%The time dependence of physical length scale of the mode $A$ 
%is $a(\eta)/k_A$, thus  the mode $A$ is created at the time when 
%the condition  $1/M_c(\eta)=a(\eta)/k_A$ is satisfied, 
%, where $a(\eta)/k_A$ is the physical length scale of the mode $A$. 
Another moment $\etaAC$ at which the mode $A$ crosses the horizon 
is again given by the equation $a(\etaAC)/k_A=1/H.$
Then the inequality $\eta^{\rm Q}_A \gtrless \eta^{\rm C}_A$ 
is translated in terms of $M_c(\eta)$ as 
\begin{align}
 \eta^{\rm Q}_A \gtrless \eta^{\rm C}_A 
  \quad \Longleftrightarrow \quad
  \frac{H}{M_c(\eta^{\rm Q}_A)} \gtrless 1, 
 \label{Mc}
\end{align}
because
\begin{align}
 \eta^{\rm Q}_A \gtrless \eta^{\rm C}_A \quad \Longleftrightarrow 
 \quad \frac{a(\eta^{\rm Q}_A)}{k_A} \gtrless 
 \frac{a(\eta^{\rm C}_A)}{k_A} \quad \Longleftrightarrow 
 \quad \frac{1}{M_c(\eta^{\rm Q}_A)}\gtrless \frac{1}{H}.
\end{align}
Here we have used the fact that $a(\eta)$ is a 
monotonically increasing function. 
Together with the discussion given in subsection 2.2, 
we thus conclude that the contribution of a mode $A$ to the CMB anisotropy 
will be suppressed largely 
if the inequality $H/M_c(\eta^{\rm Q}_A) >1$ holds strongly.  
Contrarily the contribution of a mode $A$ will be almost the same 
as in the case with the Bunch-Davies vacuum 
if the inequality $H/M_c(\eta^{\rm Q}_A) <1$ holds strongly.

How the CMB anisotropy deviates 
from a scale invariant one depends on the $\eta$ dependence of $M_c$.
If, for example, $M_c(\eta)$ is a monotonically increasing function, 
then there is a critical time $\eta_*$ such that $M_c(\eta_*) = H$, 
and we have the inequality 
$H/M_c(\eta) \gtrless 1$ for $\eta \lessgtr \eta_*$. 
%Together with Eq.\ (\ref{Mc}), 
%this implies that a mode $A$ will be supressed largely 
%if the inequality $\etaAQ<\eta_*$ is satisfied strongly. 
%Hence the contributions of the modes created at earlier times
%are suppressed.
Since larger-scale modes are created at earlier times,  
the inequality $H/M_c(\eta^{\rm Q}_A) >1$ holds more strongly 
for larger-scale modes, 
and thus their contributions to the primordial spectrum are 
more suppressed,
leading to a large-scale damping.
On the other hand, if $M_c(\eta)$ is a monotonically decreasing function, 
there will be a damping on fluctuations on small scales.
If $M_c$ is constant and $H/M_c \gg 1$ there is no CMB anisotropy, and 
if $H/M_c \ll 1$ we get an ordinary, scale invariant spectrum.

{}From these analysis, we know that $M_c$ corresponding to 
the cutoff of Fig.\ \ref{largescale} is a monotonically increasing function,
and $M_c$ corresponding to Fig.\ \ref{smallscale} 
is a monotonically decreasing function.
$M_c$ corresponding to Figs.\ \ref{nocmb} and  \ref{nodamping} 
are constant, and $H/M_c>1, H/M_c<1$, respectively.

{}For example, 
if we take the homogeneous cutoff used in subsection 2.3.1, then 
\begin{align}
 \eta^{\rm Q}_A &= - \frac{1}{k\Lcut H},
\end{align}
and thus we have
\begin{align}
\frac{1}{M_c\bigl(\eta^{\rm Q}_A\bigr)} &= 
 \frac{a\bigl(\eta^{\rm Q}_A\bigr)}{k} = \Lcut.
\end{align}
Therefore $M_c$ does not depend on $\eta$ and because we make analysis
in de Sitter spacetime, we get a scale invariant power spectrum.
However, if we take the cutoff used in subsection 2.3.2, then
\begin{align}
 \eta^{\rm Q}_{A} &= - \frac{r_*}{\Lcut H} \sqrt{ \frac{4\pi}{l+1}},
\end{align}
and thus we have
\begin{align} 
 M_c\big(\eta^{\rm Q}_{A}\big) 
 &=\left(\frac{a\big(\eta^{\rm Q}_A\big)}{k}\right)^{-1} 
 = -H \etaAQ k
 \sim 
 -\frac{2\pi^2 r_*}{\Lcut^2 H \eta^{\rm Q }_A},
\end{align}
where we have used the relation that $k \sim \pi(l+1)/2r_*$.
Therefore $M_c(\eta)$ is a monotonically increasing 
function of $\eta$, giving a large-scale damping.

%%%%%%%%%%%%%%%%%%%%%%%%%%%%%%%%%%%%%%%%%%%%%%%%%%%%%%%%%%%%%%%%
%%%%%%%%%%%%%%%%%%%%%%%%%%%%%%%%%%%%%%%%%%%%%%%%%%%%%%%%%%%%%%%%
\resection{Analysis without resorting to the classicalization process}

In the preceding section, the large-scale damping is analyzed by
comparing the two moments $\etaAQ$ and $\etaAC$ for each quantum mode $A$ 
and also by resorting to the classicalization process 
of its quantum fluctuation.
Although the discussion given there should yield a qualitatively
correct result, 
it is not satisfactory because the classicalization proceeds only
gradually around the moment when the mode crosses the horizon, and 
because the process should be automatically incorporated once the
field is quantized with a proper initial condition.
Furthermore, in the analysis made in subsection 2.3.2, 
the classical fluctuation of a mode 
is assumed to take the same magnitude with that in the absence of cutoff 
once the mode has a period of quantum fluctuation.
However, if the mode has only a short period of quantum fluctuation, 
then only a small magnitude of classical fluctuation 
will be left after the classicalization process 
(the fossil should be small if the lifetime of quantum fluctuation 
is short). 
Thus in this section, we reanalyze the angular power spectrum 
of the CMB anisotropy 
by imposing on each mode an initial condition which seems plausible. 

The initial condition we choose is such that each mode $A$ starts 
its quantum fluctuation as a vacuum fluctuation 
at the ground state of the Hamiltonian.%
\footnote{
This type of initial condition is investigated in
Ref.\ \cite{Brandenberger:2002nq} in a different context.
} 
We also assume that a cutoff is introduced in such a way that 
its physical distance scale $\Lcut$ is a time-independent constant. 
We first consider the case in which a homogeneous cutoff is introduced 
(case of subsection 2.3.1).  
Then we consider the case in which the two-sphere at $(\eta,r)$ 
is replaced by the fuzzy sphere $\tilde{S}^2$ 
(case of subsection 2.3.2). 
%%%%%%%%%%%%%%%%%%%
\subsection{Homogeneous cutoff on the physical wave length}

In this subsection, we label the mode as $A=\bm k$ 
and set the homogeneous cutoff\ (\ref{cutone}). 
We require that at the moment (\ref{etaaq}), $\etaAQ=-1/k\Lcut H$,
the mode $A=\bm k$ be at the ground state of the Hamiltonian. 

The action we consider is
\begin{align}
 S_{\rm fluct} &= \frac{1}{2} \int d\eta\, d^3\bm{x}\,  \frac{1}{H^2 \eta^2}
  \Bigl[(\partial_\eta \phi(\eta,{\bm x}))^2 - (\partial_{i}
  \phi(\eta,{\bm x}))^2 \Bigr].  
\end{align}
The conjugate momentum to $\phi$ is 
$\pi = \partial_\eta \phi/H^2 \eta^2$.
By introducing the annihilation and creation operators as 
\begin{align}
 \phi=\int \frac{d^3\bm k}{(2\pi)^3}\,
  \frac{1}{\sqrt{2 \mu_k}} (a_{\bm k} + a_{- \bm k}^\dag)
  \,e^{i{\bm k}\cdot{\bm x}}, \qquad
 \pi=-i\int \frac{d^3\bm k}{(2\pi)^3}\,
  \sqrt{\frac{\mu_k}{2}} (a_{\bm k} - a_{- \bm k}^\dag)
  \,e^{i{\bm k}\cdot{\bm x}}, 
\end{align}
\begin{align}
 \big[ a_{\bm k}, a_{\bm k'}^\dag\big] = 
  (2\pi)^3 \delta^3({\bm k}-{\bm k'}),
\end{align}
the Hamiltonian can be written as 
\begin{align}
 H &= \int \frac{d^3{\bm k }}{(2\pi)^3} \bigg[ \omega_k a_{\bm k}^\dag a_{\bm 
  k} + \frac{\lambda_k}{2}
  ( a_{\bm k}^\dag a_{-\bm k}^\dag  + a_{\bm k} a_{- \bm k} )  
  \bigg], \\
 \omega_k  &= \frac{\mu_k}{2}H^2\eta^2 + \frac{1}{2\mu_k}\frac{k^2}{H^2
  \eta^2},
  \qquad
  \lambda_k = -\frac{\mu_k}{2}H^2\eta^2 
  + \frac{1}{2\mu_k}\frac{k^2}{H^2 \eta^2}. 
\end{align}
We choose $\mu_k$ such that 
the Hamiltonian is diagonalized at the moment $\etaAQ$, 
\begin{align}
 \mu_k = \frac{k}{H^2\bigl(\etaAQ\bigr)^2}. 
\end{align}
Then our requirement that the mode $\bm k$ be at the ground state 
when $\eta=\etaAQ$ 
can be realized by the following initial condition: 
\begin{align}
 a_{\bm k}\bigl(\etaAQ\bigr) \bigl|0\bigr\rangle = 0.
\end{align}

In order to calculate the two-point correlation functions 
at the exit time of inflation, 
we take the Heisenberg picture for the annihilation-creation 
operators and consider their Bogoliubov transformations: 
\begin{align}
 \left(
  \begin{array}{c}
   a_{\bm k}(\eta) \\
   a_{- \bm k}^\dag(\eta)\\
  \end{array}
 \right)
 =
 \left(
  \begin{array}{cc}
   A_{\bm k}(\eta) & B_{\bm k}(\eta) \\
   B^\ast_{- \bm k}(\eta) & A_{- \bm k}^\ast(\eta)
  \end{array}
 \right)
 \left(
  \begin{array}{c}
   a_{\bm k}(\etaAQ)\\
   a_{- \bm k}^\dag(\etaAQ) 
  \end{array}
 \right). 
\end{align}
The coefficients $A_{\bm k}$ and $B_{- \bm k}^\ast$ 
are determined by the Heisenberg equations, 
and using the combinations 
$\xi_{\bm k}\equiv A_{\bm k}+B_{- \bm k}^\ast$ and 
$\zeta_{\bm k}\equiv A_{\bm k}-B_{- \bm k}^\ast$, 
we can express the power spectrum $P_\phi (k)$ of the inflaton field 
at the exit time of inflation ($\eta=-0$) as  
\begin{align}
 \bigl\langle 0\bigr| \phi_{\bm k}^\dag(0)\,\phi_{\bm k'}(0) 
  \bigl|0\bigr\rangle 
  \equiv \frac{2\pi^2}{k^3}P_\phi(k)\, (2\pi)^3
  \delta^3({\bm k} - {\bm k'})
  = \frac{1}{2\mu_k}\big| \xi_{\bm k}(0) 
  \big|^2 (2\pi)^3\delta^3({\bm k} - {\bm k'}), 
\end{align}
or
\begin{align}
  P_\phi (k) = \frac{k^3}{(2\pi)^2 \mu_k} \big| \xi_{\bm k}(0) \big|^2.
\end{align}
By using the identity, 
\begin{align}
  e^{i {\bm k}{\bm x}} = 4\pi \sum_{lm}i^l
  j_l(kr)Y_{lm}^\ast(\Omega_{\bm k})Y_{lm}(\Omega),
\end{align}
$P_\phi(k)$ can be related to the 
angular power spectrum $C_l^\phi(0,r_*)$ as
\begin{align}
  C_l^\phi(0,r_*) = 4\pi \int_0^\infty
  \frac{dk}{k} \big( j_l(kr_*) \big)^2 P_\phi(k).
\end{align}
Note that a scale invariant $P_\phi(k)$ leads to $C_l^\phi(0,r_*)
\propto 1/l(l+1)$.
The Heisenberg equations give the following equations :
\begin{align}
  \ddot{\xi}_{\bm k}(\eta) -\frac{2}{\eta}\dot{\xi}_{\bm k}(\eta) 
  + k^2\xi_{\bm k}(\eta)  = 0, \qquad  \xi_{\bm k}(\etaAQ) =1, \qquad
  \dot{\xi}_{\bm k}(\etaAQ) = -ik. 
\end{align}
They can be solved with the spherical Hankel function as 
\begin{align}
 \xi_{\bm k}(\eta) &= \frac{i}{2}
  \bigg(\frac{\eta}{\etaAQ}\bigg)^{\!2}\Bigl( \bigl(1-2ik\etaAQ\bigr)
   e^{ik\etaAQ}\,
  h_1^{(1)}(-k\eta) + e^{-ik\etaAQ}\, h_1^{(2)}(-k\eta)  \Bigr),
\end{align}
and thus, the power spectrum is determined as 
\begin{align}
 P_\phi(k) =
  \bigg(\frac{H}{2\pi}\bigg)^2\,\frac{1}{4\big(k {\etaAQ}\big)^2}\,
  \Big|\,1-2ik {\etaAQ}-\exp(-2ik\etaAQ)  \,\Big|^2. 
\end{align}
Substituting the equation $\eta_A^{\rm Q}= -1/k L_{\rm cut} H$, we get 
\begin{align}
 P_{\phi}(k) \Big|_{\etaAQ=-1/k \Lcut H} 
  &= \biggl(\frac{H}{2\pi}\biggr)^2\,\frac{\bigl(\Lcut H\bigr)^2}{4}\,
  \bigg|\, 1 +\frac{2i}{\Lcut H} 
  - \exp\left( \frac{2i}{\Lcut H} \right)\,\bigg|^2 \notag \\
 \Bigl(&= k \,{\rm -independent}\Bigr). \label{Pphi2}
\end{align}
Thus we obtain a scale invariant power spectrum for the homogeneous
cutoff in our approximation that the background metric is a pure de
Sitter universe. This result is consistent with a qualitative
argument given in subsection 2.3.1.
%In subsection 2.3.1 and here we calculate the power spectrum of 
%inflaton field in de Sitter space with homogeneous cutoff.
%In the former we discuss using the relation 
%between the two moments $\etaAQ$ and $\etaAC$, 
%and in the latter we calculate setting the initial condition for each mode 
%at the moment $\etaAQ$.
%In both cases we get neither large-scale nor small-scale damping 
%consistently.
%\footnote{
%We get the exactly scale invariant power spectrum,
%because we analyze the power spectrum
%in de Sitter space.
%This analysis corresponds to the zeroth order of 
%the slow roll approximation. 
%If we consider corrections we will get 
%non scale-invariant result. 
%We neglect these corrections 
%because we are interested in the mechanism with which 
%we get the large-scale damping in the zeroth order of the 
%slow roll approximation.
%}  
If we take the limits 
$\Lcut H \to 0$ and $\Lcut H \to \infty$ in Eq.\ \eqref{Pphi2} we get
\begin{align}
  P_\phi \big|_{\Lcut H \to 0}
  &=
  \left( \frac{H}{2 \pi} \right)^2, \\
  P_\phi \big|_{\Lcut H \to \infty}
  &=
  0.
\end{align}
This is an expected behavior for a proper initial condition,
as was discussed in subsection 2.2.

%%%%%%%%%%%%%%%%% 
%%%%%%%%%%%%%%%%%
\subsection{Cutoff with fuzzy sphere $\tilde{S}^2$} 

In this subsection, we consider the model considered 
in subsection 2.3.2, 
in which three-dimensional space is described by 
$\bR_+ \times \tilde{S}^2$ 
(radial coordinate $r$ times a fuzzy sphere). 
The noncommutativity is introduced only to the angular directions 
$\Omega$, since this would be most relevant to any modifications 
to angular power spectrum.%
\footnote{ 
In fact, recently it has been shown in 
Refs.\ \cite{Tsujikawa:2003gh,Huang:2003hw} 
that there is no sharp damping at large angular scales 
if one introduces noncommutativity only to 
the time and radial coordinates.
}
One can also give a holographic interpretation 
to this way of introducing cutoff, 
which is briefly commented in conclusion 
and will be discussed more elaborately in a forthcoming paper \cite{fkm3}. 

%%%%%%%%%
\subsubsection{Coordinates}

\begin{figure}[htbp]
 \begin{center}
  \includegraphics[width=7cm]{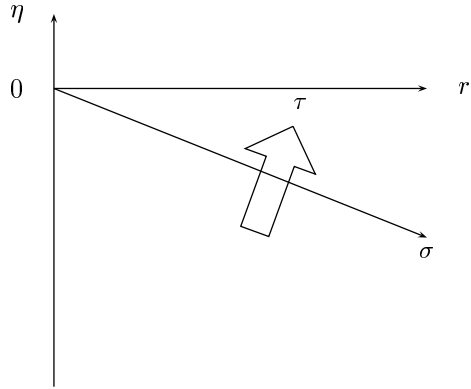}
 \end{center}
 \caption{\footnotesize New coordinates $(\tau,\sigma)$. 
   The moment when the mode $A=(p,l,m)$ newly appears 
   is expressed by a single time-slice. }
\label{cord}
\end{figure}

As was explained in subsection 2.3.2, 
the mode with the multipole $l$ 
starts its quantum fluctuation 
at the moment (\ref{etal}) 
which depends also on the value of the radial coordinate $r$. 
Thus, we introduce the new coordinates $(\tau,\sigma,\Omega)$ 
such that the modes with the same $l$ start at the same time: 
\begin{align}
 H\eta = -\,e^\sigma f_1(\tau), \quad 
  Hr = e^\sigma f_2(\tau).
\end{align}
We here assume that $f_1,\,f_2>0$ 
and that $\partial_\tau (f_2/f_1)>0$ 
so that the orientations of $\eta$ and $\tau$ coincide.%
\footnote{
Recall that $\eta<0$ during inflation.
} 
Note that 
\begin{align}
 \partial_\sigma = \eta\,\partial_\eta + r\,\partial_r 
\end{align}
is the Killing vector of the de Sitter metric, 
so that the metric does not depend on $\sigma$.
Setting $p$ to be the conjugate momentum to the coordinate $\sigma$, 
we label the mode as $A=(p,l,m)$.   
The cutoff is set on the multipole $l$ as in Eq.\ (\ref{lmax}) 
and we denote by $\tau = \tau_A^{\rm Q}$ the moment 
when the mode $A=(p,l,m)$ starts to fluctuate quantum mechanically 
(see Fig.\ \ref{cord}):
\begin{align}
 \frac{f_2}{f_1} \bigg|_{\tau=\tau_A^{\rm Q}} = \Lcut H
 \sqrt{\frac{l+1}{4 \pi}}\quad (= \alpha_l).
\end{align}

%%%%%%%%%%%%%%
%%%%%%%%%%%%%%%%

%%%%%%%%%%
%%%%%%%%%%
\subsubsection{Calculation of the angular power spectrum}

We now calculate the angular power spectrum.  
We will see that it is independent of the explicit form 
of the functions $f_1$ and $f_2$. 

Under the above parametrization, the metric is rewritten as
\begin{align}
 ds^2 &= \frac{1}{H^2\eta^2}\,\bigl(-d\eta^2+dr^2+r^2d\Omega^2\bigr)\nonumber\\
 &= \frac{1}{H^2 f_1^2}\Big(\, \big(\dot{f}_2^2 - \dot{f}_1^2\big)
  \,d\tau^2 
  + \big(f_2^2- f_1^2   \big)\,d\sigma^2 
  + 2\big( f_2 \dot{f}_2 - f_1 \dot{f}_1 \big)\,d\tau d\sigma 
  + f_2^2\,d\Omega^2 \,\Big).
\end{align}
Here the dot stands for the differentiation with respect to $\tau$. 
The action can be expanded in the angular modes $(l,m)$. 
 Due to the rotational invariance, 
it is enough to consider the case $m=0$:  
\begin{align}
 S_{{\rm fluct} (l,0)} & = \int d\tau d\sigma \frac{f_2^2}{2H^2f_1^2 F}
 \times \notag \\
  &\hspace{-1.5cm}\times \bigg[(f_2^2-f_1^2) (\partial_\tau \phi_{l0} )^2 +
  (\dot{f}_2^2 -\dot{f}_1^2) (\partial_\sigma \phi_{l0})^2 - 2(f_2
  \dot{f}_2 -f_1 \dot{f}_1) 
  \partial_\tau \phi_{l0} \partial_\sigma \phi_{l0} 
  - \frac{F^2 l(l+1)}{f_2^2}\phi_{l0}^2\bigg], 
\end{align}
where $F \equiv f_1 \dot{f}_2 - f_2 \dot{f}_1 
= f_1^2\,\partial_\tau(f_2/f_1)>0$. 
The conjugate momentum to $\phi_{l0}$ is 
\begin{align}
 \pi_{l0}%=\frac{\delta S_{l0}}{\delta \dot{\phi}_{l0}} 
  = \frac{f_2^2}{H^2
  f_1^2 F}\big[ (f_2^2 -f_1^2)\partial_\tau \phi_{l0} -(f_2
  \dot{f}_2 - f_1 \dot{f}_1) \partial_\sigma \phi_{l0}  \big]. 
\end{align}
By introducing the annihilation and creation operators as 
\begin{align}
 &\phi_{l0} = \int_{- \infty}^\infty
  \frac{dp}{2\pi} \frac{1}{\sqrt{2 \mu_p}} (a_p +
  a_{-p}^\dag) e^{ip\sigma} , \qquad
 \pi_{l0} = -i \int_{- \infty}^\infty
  \frac{dp}{2\pi} \sqrt{\frac{\mu_p}{2}}(a_p -
  a_{-p}^\dag) e^{ip\sigma} ,
\end{align}
\begin{align}
  \big[ a_p, a_{p'}^\dag\big] = 2\pi\,\delta(p-p'),
\end{align}
the Hamiltonian is given by 
\begin{align}
 H &= \int_{-\infty}^\infty
  \frac{dp}{2\pi} \left( \omega_p a_p^\dag a_p +
   \frac{\lambda_p}{2}(a_p^\dag a_{-p}^\dag + a_p a_{-p})   \right) 
\end{align}
with
\begin{align}
 \omega_p & \equiv \frac{\mu_p H^2 f_1^2 F}{2 f_2^2(f_2^2 - f_1^2)} +
  \frac{f_2^2 F}{2\mu_p H^2 f_1^2} \left( \frac{p^2}{f_2^2-f_1^2} +
  \frac{l(l+1)}{f_2^2}  \right) -p\frac{f_2 \dot{f}_2 - f_1
  \dot{f}_1}{f_2^2-f_1^2}, \\
 \lambda_p & \equiv -\frac{\mu_p H^2 f_1^2 F}{2 f_2^2(f_2^2 - f_1^2)} +
  \frac{f_2^2 F}{2\mu_p H^2 f_1^2} \left( \frac{p^2}{f_2^2-f_1^2} +
  \frac{l(l+1)}{f_2^2}  \right). 
\end{align}
As in the previous subsection, 
we set $\mu_p$ such that the Hamiltonian is diagonalized 
at $\tau = \tau_A^{\rm Q}$, making  
the mode to start its vacuum fluctuation at the ground state:  
\begin{align}
 &\mu_p = \frac{f_2^2}{H^2 f_1^2} \sqrt{p^2 + \frac{f_2^2 -f_1^2}{f_2^2}
  l(l+1)} \Bigg|_{\tau = \tau_A^{\rm Q}}, \\
 &a_p \big(\tau_A^{\rm Q}\big) \,\big|0 \big\rangle =0.
\end{align}
We again consider the Bogoliubov transformations:
\begin{align}
 \left(
  \begin{array}{c}
   a_p(\tau) \\
   a_{- p}^\dag(\tau)\\
  \end{array}
 \right)
 =
 \left(
  \begin{array}{cc}
   A_p(\tau) & B_p(\tau) \\
   B^\ast_{- p}(\tau) & A_{- p}^\ast(\tau)
  \end{array}
 \right)
 \left(
  \begin{array}{c}
   a_p(\tau_A^{\rm Q})\\
   a_{- p}^\dag(\tau_A^{\rm Q}) 
  \end{array}
 \right), 
\end{align}
and introduce $\xi_p\equiv A_p + B_{-p}^\ast$ and 
$\zeta_p \equiv A_p - B_{-p}^\ast$,  
which satisfy the differential equations
\begin{align}
 \partial_\tau \xi_p &= 
  -\frac{i}{2}\,\big(\omega_p - \omega_{-p}\big)\, \xi_p 
  -\frac{i}{2}\,\big(\omega_p + \omega_{-p}-2\lambda_p\big)\,\zeta_p, \\
 \partial_\tau \zeta_p &= 
  -\frac{i}{2}\,\big(\omega_p + \omega_{-p} +2 \lambda_p\big)\, \xi_p
  -\frac{i}{2}\,\big(\omega_p - \omega_{-p}\big)\, \zeta_p ,
\end{align}
with the initial conditions
\begin{align}
 \xi_p\big(\tau_A^{\rm Q}\big) = \zeta_p\big(\tau_A^{\rm Q}\big) = 1. 
\end{align}
The angular power spectrum is then expressed as 
\begin{align}
 l(l+1)C_l^\phi = l(l+1)\big\langle0\big| \phi_{l0} (\tau,\sigma)\,
  \phi_{l0}(\tau,\sigma) \big|0 \big\rangle \Big|_{\eta = -0} 
  = \int_0^\infty \frac{dp}{2\pi}\,\frac{l(l+1)}{\mu_p} \,
  \big|\xi_p(\tau)\big|^2  \Big|_{\eta = -0} \label{l(l+1)Cl}. 
\end{align}

The time-evolution of $\big|\xi_p(\tau)\big|^2$ can be evaluated numerically. 
This can be carried out easily if one notes that the functions  
\begin{align}
 K_p &\equiv \frac{l(l+1)}{\mu_p H^2}\big|\xi_p(\tau)\big|^2, \\
 L_p &\equiv \frac{i}{2}\,l(l+1)\bigl(\xi_p^\ast(\tau) \zeta_p(\tau) 
  - \xi_p(\tau)\zeta_p^\ast(\tau)\bigr), \\
 M_p &\equiv l(l+1)\mu_p H^2\big|\zeta_p(\tau)\big|^2, 
\end{align}
satisfy the following first-order equations: 
\begin{align}
 \partial_t K_p &= -\frac{2}{t^2(t^2 - 1)} L_p, 
  \label{Keqt}\\
 \partial_t L_p &= \left( \frac{t^2 p^2}{t^2-1} + l(l+1)\right)
  K_p - \frac{1}{t^2(t^2-1)} M_p, \label{Leqt}\\
 \partial_t M_p &= 2 \left( \frac{t^2 p^2}{t^2-1} + l(l+1)
  \right) L_p, \label{Meqt} 
\end{align}
with the initial conditions
\begin{align}
 K_p\big(t_A^{\rm Q}\big) = \frac{l(l+1)}{\mu_p H^2},
  \qquad L_p\big(t_A^{\rm Q}\big) =0,
  \qquad M_p\big(t_A^{\rm Q}\big) = l(l+1)\mu_p H^2 \label{ic},
\end{align}
\begin{align}
 \mu_p H^2= t_A^{\rm Q} \sqrt{\big(t_A^{\rm Q}\big)^2 p^2 
  + \Big(\big(t_A^{\rm Q}\big)^2-1\Big)l(l+1)} \label{mu}. 
\end{align}
Here we have introduced a new variable $t$ as
\begin{align}
 t \equiv -\frac{r}{\eta} = \frac{f_2}{f_1} , \label{newt}
\end{align}
which takes the following value at the initial time: 
\begin{align}
  t_A^{\rm Q} \equiv -\frac{r}{\etaAQ}=
  \frac{f_2\big(\tau_A^{\rm Q}\big)}{f_1\big(\tau_A^{\rm Q}\big)} 
  = \Lcut H \sqrt{\frac{l+1}{4\pi}}.
 \label{init_time}
\end{align}
We also have used the relation  
\begin{align}
 \frac{F}{f_1^2} = \partial_\tau t. 
\end{align}
The exit time of inflation  $(\eta \to -0)$ corresponds to 
the limit $t \to +\infty$. We thus have 
\begin{align}
 l(l+1) C_l^\phi = H^2 \int^\infty_0 \frac{dp}{2\pi}
 K_p(t) \Big|_{t=+\infty}. \label{Clphi_t}
\end{align} 
Since all the expressions (\ref{Keqt})--(\ref{mu}),
(\ref{init_time}) and (\ref{Clphi_t}) 
are written in terms of $t$,
the angular power spectrum $C_l^\phi$ is actually independent of the
choices of $f_1$ and $f_2$.

\begin{figure}[htbp]
 \begin{center}
  \includegraphics[width=7cm,angle=-90]{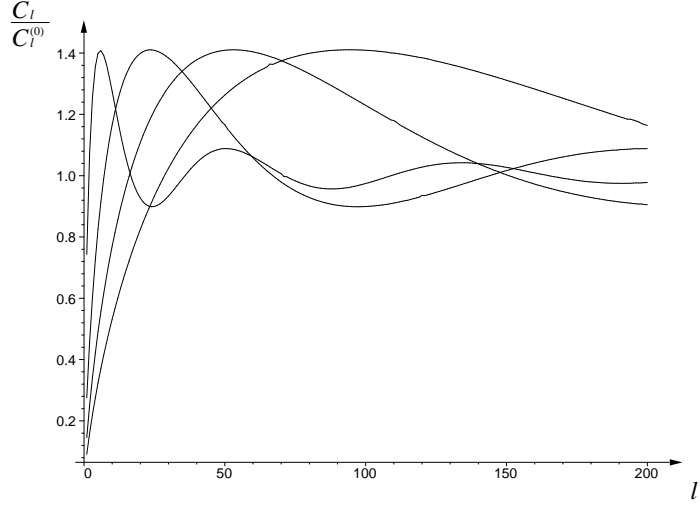}
 \end{center}
 \caption{\footnotesize 
  Damping factor for 
  the space $R_+ \times \tilde{S}^2$. 
  The horizontal axis represents the multipole $l$, 
  while the vertical axis is $C_l/C_l^{(0)} = 2\pi l(l+1)C_l^\phi /H^2$. 
  $C_l^{(0)}$ represents the angular power spectrum 
  in the absence of noncommutativity. 
  Here $\Lcut H=5, 10, 15, 20$ from top to bottom. 
  All curves obey the approximate scaling law explained in the main text, 
  and converges to one for large $l$. }
\label{NC_CMB}
\end{figure}

We carried out a numerical calculation  
and have obtained the angular power spectrum 
given in Fig.\ \ref{NC_CMB}. 
\begin{figure}[htbp]
 \begin{center}
  \includegraphics[width=7cm,angle=-90]{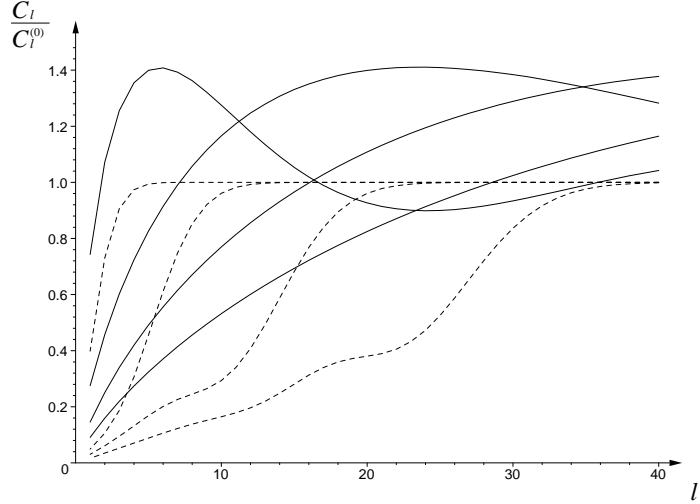}
 \end{center}
 \caption{\footnotesize 
  Damping factor for 
  the space $R_+ \times \tilde{S}^2$. 
  The horizontal axis represents the multipole $l$, 
  while the vertical axis is $C_l/C_l^{(0)}$.
  Dashed lines are the spectra  
  obtained with the ansatz \eqref{ansatz}, 
  while solid lines are
  the spectra given in this section. 
  In each set $\Lcut H = 5,10,15,20$ from top to bottom.
  Here the spectral index is set to be $n=1$ for both.
}
\label{NC_CMB2}
\end{figure}%
Figure \ref{NC_CMB2} is a magnified figure for small $l$,  
where the power spectra obtained with the ansatz 
\eqref{ansatz} are also drawn for comparison 
(the spectral index is set to be one for both).  
%together with part of Fig.\ \ref{NC_CMB} for comparison.
%the results given in subsection 2.3.2 is also depicted there.
The spectra in Fig.\ \ref{NC_CMB} 
are oscillating with $l$ and approach one asymptotically.
This oscillation is a common feature of the power spectra 
when taking initial conditions other than the Bunch-Davies vacuum.%
\footnote
{
See Refs.\ \cite{Martin:2000xs,Cremonini:2003yv,Easther:2001fz}.
Note that we had no such oscillation in subsection 3.1. 
It is because we there analyzed the power spectrum with the use of 
the homogeneous cutoff in de Sitter spacetime.   
} 
We also see that the resulting power spectra 
exhibit less sharp damping than the ones obtained with the ansatz \eqref{ansatz}.
Although the damping we get here may not be sufficiently large 
to fit the observational data, 
the curve could be better fit to the data 
by taking account of corrections in the slow roll approximation 
(i.e., including a nontrivial spectral index), 
or it is still possible that our initial condition needs to be modified.%
\footnote{
Of course, another way to understand the result is to consider 
the cosmic variance (i.e., to regard the observational data 
as a deviation from the obtained result above). 
}

%%%%%%%
\subsubsection{Scaling law in the angular power spectrum}

As can be seen from Fig.\ \ref{NC_CMB}, 
the curves of the angular power spectrum shift to right 
for larger $\Lcut H$. 
In fact, one can prove that the angular power spectrum 
is a function of the scaling variable $l/(\Lcut H)^2$ 
with good precision. 
In order to show this, 
we assume that $ t_A^{\rm Q} \gg 1 $ in Eqs.\ (\ref{Keqt})--(\ref{mu}),%
\footnote{
As can be seen from Eq.\ (\ref{init_time}), 
this assumption holds well when  $\Lcut H$ and/or $l$ are large. 
}
and rescale $p$ and $t$ in Eq.\ (\ref{Clphi_t})
such that
\begin{align}
 p = k \sqrt{l(l+1)} , \qquad
  t = s\,t_A^{\rm Q} \quad\Bigl(t=t_A^{\rm Q}\Leftrightarrow s=1
  \Bigr).
\end{align}
Then by setting
\begin{align}
 {\cal K}_k \equiv \sqrt{l(l+1)}K_p ,\qquad
  {\cal L}_k \equiv \frac{\sqrt{l(l+1)}L_p}{\big(t_A^{\rm Q}\big)^3},
  \qquad
  {\cal M}_k \equiv \frac{\sqrt{l(l+1)}M_p}{\big(t_A^{\rm Q}\big)^6}, 
\end{align}
all the equations depend only on the combination 
$l/(\Lcut H)^2= l(l+1)/4\pi \bigl(t_A^{\rm Q}\bigr)^2$ and on $k$:
\begin{align}
 \partial_s {\cal K}_k &= -\frac{2}{s^4} {\cal L}_k,\\
 \partial_s {\cal L}_k &= (k^2+1)\frac{l(l+1)}{\big(t_A^{\rm Q}\big)^2} 
  {\cal K}_k - \frac{1}{s^4} {\cal M}_k,\\
 \partial_s {\cal M}_k &= 2(k^2+1) \frac{l(l+1)}{\big(t_A^{\rm Q}\big)^2} 
  {\cal L}_k,
\end{align}
with
\begin{align}
 {\cal K}_k(1) = \frac{l(l+1)}{\big(t_A^{\rm Q}\big)^2\sqrt{(k^2+1)}} ,
  \quad
 {\cal L}_k(1) =0 , 
  \qquad 
 {\cal M}_k(1) = \sqrt{k^2+1} \frac{(l(l+1))^2}{\big(t_A^{\rm Q}\big)^4}. 
\end{align}
Since the angular power spectrum is expressed as the integration of 
${\cal K}_k$ over $k$:
\begin{align}
 l(l+1)C_l^\phi = H^2 \int_0^\infty \frac{dp}{2\pi}\,K_p(t) \Big|_{t=+\infty}
   = H^2\int_0^\infty \frac{dk}{2\pi} \,{\cal K}_k(s) \Big|_{s=+\infty},
\end{align}
we conclude that the angular power spectrum 
is a function only of $l/(\Lcut H)^2$ 
when one can assume that  $ t_A^{\rm Q} \gg 1 $.

%%%%%%%%%%%%%%%%%%%%%%%%%%%%%%%%%%%%%%%%%%%%%%%%%%%%%%%%%%%%%%%%
%%%%%%%%%%%%%%%%%%%%%%%%%%%%%%%%%%%%%%%%%%%%%%%%%%%%%%%%%%%%%%%%
\resection{Conclusion and outlook}

In this article, we have investigated 
the effects of cutoff or noncommutativity 
on the CMB anisotropy, 
focusing on the possibility that the large-scale damping 
observed in the CMB angular power spectrum  
may be explained as such effect. 

We gave an analysis in two ways: 
the first was based on the classicalization process, 
and the second was carried out by analyzing the time-evolution 
of the Heisenberg operators with the initial condition 
such that modes start as vacuum fluctuations.

In the first analysis (section 2), 
we have clarified the mechanism 
that a short-distance cutoff can affect the CMB anisotropies 
on large scales. 
We there introduced two typical moments during inflation, 
one is the moment $\etaAC$ 
when a given mode $A$ crosses the Hubble horizon and becomes classical, 
and the other is the moment $\etaAQ$ at which 
the mode $A$ starts to fluctuate quantum mechanically. 
We have shown that there exists a large-scale damping 
if $\etaAC$ is prior to $\etaAQ$ only for larger-scale modes. 
{}Furthermore, we have demonstrated that the homogeneous 
three-dimensional cutoff does not yield such damping, 
while the assumption that space is the product of 
the radial coordinate and the fuzzy sphere with 
%time-independent 
short-distance cutoff $\Lcut$ 
%(in physical scale), 
certainly exhibits the large-scale damping 
if one assumes that the noncommutative scale is 
around the Hubble parameter during inflation.%
\footnote{
In Ref.\ \cite{Fukuma:2003kv}, a cyclic model of the universe is considered,
giving the scenario that there is an era during which the universe
expands exponentially with the Hubble parameter equal to the string
mass scale.
}

In the second analysis based on the Heisenberg equations (section 3), 
we again have shown that the homogeneous three-dimensional cutoff 
only gives rise to the scale invariant power spectrum, 
while the cutoff based on the fuzzy sphere 
certainly yields the power spectrum which exhibits 
the large-scale damping.

We thus may conclude that our analysis presented in this article 
gives a qualitatively correct explanation 
on the mechanism of the large-scale damping. 
However, the figure does not exhibit enough damping 
compared to the observational data.
This would be improved if we introduce a nontrivial spectral index.
Another possible way around is to modify our initial condition. 
In fact, one would demand that when a mode newly appears 
its amplitude should be as small as possible, 
but the vacuum fluctuation may not be the least 
amount of fluctuation among those that are allowed. 
It would be nice if one could find a proper initial condition 
which correctly describes the creation of modes 
after being released from the constraint of cutoff. 
The initial conditions in chaotic inflation \cite{Linde:gd} 
or in the models of ``the universe from nothing'' \cite{Vilenkin:de} 
would be interesting to investigate in this context.%
\footnote
{
In searching for desirable initial conditions, 
one will need to carefully check whether the choice of initial condition 
avoids the backreaction problem
(see for example Refs.\ \cite{Tanaka:2000jw,Kaloper:2002cs}).
In fact, some initial conditions can cause particle creation, 
and the energy density of the particles 
could become comparable to the background energy density 
which was first assumed to be the main source driving inflation. 
}

Another point which should be clarified in the future 
is the fact that the homogeneous three-dimensional cutoff 
does not lead to a large-scale damping  
while the cutoff based on fuzzy sphere actually does. 
This may not be satisfactory from the viewpoint of 
the standard cosmology 
in the sense that the latter way of cutoff does not respect 
the cosmic principle or the invariance 
under the three-dimensional Euclidean group $\big(E(3)=ISO(3)\big)$. 
However, from the holographic viewpoint
\cite{'tHooft:gx}, 
this way of cutoff with the fuzzy sphere 
would be consistent with the following scenario of inflation. 
{}First, since the noncommutative scale in our model is close to the 
Hubble parameter during inflation, 
matter fields are strongly coupled to gravitation, 
so that the holographic nature would dominate the dynamics 
of the system. 
We further assume that the holographic principle holds 
for arbitrary three-dimensional spacelike region 
(not only for lightsheets \cite{Bousso:1999xy})
if one considers a combined system of matter fields and gravity. 
Then the dynamical degrees of freedom in any three-dimensional 
region should be controlled by the information on the boundary. 
If the inflaton (or any other scalar-like collective mode 
in the system) represents such degrees of freedom, 
then the holographic cutoff for any three-dimensional ball 
would be to set a noncommutativity on the spherical boundary. 
This line of investigation is now in progress 
and will be reported elsewhere.

%%%%%%%%%%%%%%%%%%%%%%%%%%%%%%%%%%%%%%%%%%%%%%%%%%%%%%%%%
%%%%%%%%%%%%%%%%%%%%%%%%%%%%%%%%%%%%%%%%%%%%%%%%%%%%%%%%%
\section*{Acknowledgments}
The authors would like to thank P.\ Freund, T.\ Kobayashi and  
S.-H.\ Tye for useful discussions. 
This work is supported in part by Grant-in-Aid (No.\ 15540269) 
from the Japan Ministry of Education, Culture, Sports, 
Science and Technology, 
and by Grant-in-Aid for the 21st Century COE 
``Center for Diversity and Universality in Physics." 

%%%%%%%%%%%%%%%%%%%%%%%%%%%%%%%%%%%%%%%%%%%%%%%%%%%%%%%%%
%%%%%%%%%%%%%%%%%%%%%%%%%%%%%%%%%%%%%%%%%%%%%%%%%%%%%%%%%

\end{document}